\documentclass[12pt]{amsart}
\usepackage{amsmath}
\usepackage{amssymb}
\usepackage[latin1]{inputenc}
\usepackage[T1]{fontenc}
\newtheorem{lemma}{Lemma}

\newtheorem{proposition}{Proposition}
\newtheorem{definition}{Definition}
\newtheorem{remark}{Remark}
\newtheorem{corollary}{Corollary}

\newcommand{\tr}{{\rm Tr }}

\newcommand{\bra}{\langle}
\newcommand{\ket}{\rangle}
\newcommand{\vp}{\varphi}

\newcommand{\C}{\mathbb{C}}
\newcommand{\N}{\mathbb{N}}

\newcommand{\R}{\mathbb{R}}
\newcommand{\Z}{\mathbb{Z}}
\newcommand{\be}{\begin{equation}}
\newcommand{\eeq}{\end{equation}}
\newcommand{\bet}{\begin{equation*}}
\newcommand{\eeqt}{\end{equation*}}
\newcommand{\bea}{\begin{eqnarray}}
\newcommand{\eeqa}{\end{eqnarray}}
\newcommand{\beat}{\begin{eqnarray*}}
\newcommand{\eeqat}{\end{eqnarray*}}
\newcommand{\goesto}{\rightarrow}
\newcommand{\goto}{\goesto}
\newcommand{\h}[1]{\mathcal{#1}}
\newcommand{\hil}{\mathcal{H}}

\newcommand{\hB}{\mathcal{B}}

\newcommand{\cc}[1]{\overline{#1}}

\newcommand{\lin}{{\rm lin} \,}
\newcommand{\Dcoh}{D_{coh}^2}

\begin{document}
\title{On the moment limit of quantum observables, with an application to the balanced homodyne detection}
\author{J. Kiukas}
\address{Jukka Kiukas,
Department of Physics, University of Turku,
FIN-20014 Turku, Finland}
\email{jukka.kiukas@utu.fi}
\author{P. Lahti}
\address{Pekka Lahti,
Department of Physics, University of Turku,
FIN-20014 Turku, Finland}
\email{pekka.lahti@utu.fi}

\begin{abstract}
We consider the moment operators of the observable (i.e. a semispectral measure or POM) associated with the
balanced homodyne detection statistics, with paying attention to the correct domains of these unbounded operators.
We show that the high amplitude limit, when performed on the moment operators, actually determines uniquely the entire statistics of
a rotated quadrature amplitude of the signal field, thereby verifying the usual assumption that the homodyne detection
achieves a measurement of that observable.
We also consider, in a general setting, the possibility of
constructing a measurement of a single quantum observable from a sequence of observables by taking the limit on the level of
moment operators of these observables. In this context, we show that under some natural conditions (each of which is
satisfied by the homodyne detector example), the existence of the moment limits ensures that the underlying probability measures
converge weakly to the probability measure of the limiting observable.
The moment approach naturally requires that the observables be determined by their moment operator sequences (which
does not automatically happen), and it turns out, in particular, that this is the case for the balanced homodyne detector.
\end{abstract}

\maketitle

\section{Introduction}

The balanced homodyne detection has long been an important tool in quantum optics, because it is assumed to provide a way to
measure the rotated quadrature amplitudes $\frac 1 {\sqrt 2} (e^{-i\theta} a+e^{i\theta} a^*)$
of a single mode electromagnetic field.
(See \cite{Yuen, YuenII, Yurke, Braunstein, DAriano, Banaszek} for theoretical considerations.)
An important application of this technique is
the quantum state estimation, i.e. quantum tomography (See \cite{Paris} for a collection of articles concerning this topic.)
The significance of the homodyne measurement in that context is due to the fact that the totality of the rotated
quadratures
constitute an informationally complete set of observables, so that a quantum state is determined by their combined
measurement statistics.
The balanced homodyne detection also plays a role in e.g. continuous variable quantum teleportation \cite{BraunsteinII}.

The measurement scheme is the following: A signal field is
mixed with an auxiliary field by means of a 50-50 beam splitter (possibly followed by a phase shifter), and the difference of photon
numbers on the output ports of the splitter
is detected. The auxiliary field is taken to be an oscillator in a coherent state, operating with the same frequency as
the signal field. When the strength of the auxiliary field is high, the above mentioned photon difference statistics are
considered to resemble those of a rotated field quadrature, the rotation angle being identified as the fixed phase difference
between the input signal and auxiliary fields.

The simplest description of this phenomenon is the following: Assuming that the beam splitter is lossless and causes no phase shift between the input modes,
the process is described by a unitary
operator $U$ (see \cite{Scully}), transforming the field annihilation operators $a$ (signal) and $b$ (auxiliary) into
$\tilde a = U^*aU = \frac 1 {\sqrt 2}(a- b)$, and
$\tilde b=U^*bU = \frac 1 {\sqrt 2}(a+b)$. With respect to the initial state of the two-mode field
(Heisenberg picture), the photon number difference observable is then
$\tilde b^*\tilde b-\tilde a^*\tilde a =ab^*+a^*b$, provided that the photon detectors are ideal.
(Of course, these formal operator relations are made precise by restricting the operators to a suitable dense subspace of the tensor product Hilbert space associated with the two mode system.)

According to the heuristic explanation, the strong auxiliary field can be treated ''classically'', by replacing
$b$ with the complex field amplitude $\beta =re^{i\theta}$ in the above operator, with $\theta$ identified as the phase relative to the input signal field.
Consequently, by suitably scaling the resulting operator with a factor proportional to the
intensity of the auxiliary field, one recovers the rotated
quadrature $\frac 1 {\sqrt 2} (e^{-i\theta} a +e^{i\theta}a^*)$.

When the auxiliary field is treated according to quantum mechanics, the high amplitude limit must be considered more
carefully. This kind of treatment was already given in \cite{YuenII}, where the balanced homodyne detection scheme
was first proposed. In that paper, however, as well as in \cite{Yurke}, only the first and second moments of
the photon difference statistics were calculated before taking the limit.
As is well known, these do not necessarily determine the entire statistics.
The limit procedure, as well as the photon difference statistics for finite amplitude, were examined more carefully
in \cite{Braunstein} and \cite{DAriano}, by considering the asymptotic behavior
of the characteristic functions of the associated probability measures using formal operator
series expansions. In \cite{Vogel}, the
scheme was described in terms of positive operator measures, and the characteristic functions for the probability measures
corresponding to the coherent input states of the signal field were calculated.
The high amplitude limit for these functions was then shown to
be equal to the characteristic functions for the corresponding statistics of the rotated quadratures, and the Levy-Cramer theorem
was applied to prove that the associated probability measures actually converge weakly, in the sense defined in \cite[p. 11]{Billingsley}.
(The same theorem had been previously applied to the unbalanced, homodyne detector, with a similar
conclusion \cite{Grabowski}.) However, in the calculations leading to the above mentioned characteristic functions,
some of the problematic matematical questions (like term-by-term integration of a series) were not explicitly discussed, making it
difficult to verify whether the treatment was entirely rigorous.

In this paper, we consider the problem of the
high amplitude limit in view of the moment operators $L(x^k,E)$ of the observables (i.e. semispectral
measures) $E$ arising from the homodyne detection statistics. (The moment operator
$L(x^k,E)$ is basically the ''weak'' operator integral of the power $x^k$ over $E$; the precise
definition of this unbounded operator will be given in the next section.)
The moment approach was used also in \cite{Banaszek},
where the authors studied the moment operators of the observable obtained when taking into account the imperfectness of the photon detectors.
They viewed the moment operator sequence as the collection of ''operational observables'' describing the effects caused by the measuring
arrangement, as opposed to the ''intrinsic'' observable, the rotated quadrature, which is the observable intended to be measured.
However, they dismiss the limit problem by applying the heuristic classical approximation of the auxiliary field to the
characteristic functions, so that their results do not contribute to our problem.

We begin our approach in a more general setting: We take a sequence $(E^n)_{n\in \N}$ of measurements (that is, observables), 
with the property that for each $k\in \N$ and (a suitable) vector state $\vp$, the numerical sequence
$(M_{n,\vp}^k)_{n\in \N}$ converges, where $M_{n,\vp}^k\in \R$ is the $k$th moment of the outcome statistics obtained by measuring $E^n$ in the state $\vp$. (Note that these moments are the averages
of the moment operators of the corresponding observables.)
Then we consider the question under what conditions there exists a unique limiting observable $E$, 
such that the moments $M^k_\vp$ of the measurement outcome statistics of $E$ in the state $\vp$
would be obtained as $M^k_\vp = \lim_{n\goesto \infty} M^k_{n,\vp}$.
Also, we study the implications of this convergence on the level of the effect operators $E^n(B)$, $B\in \h B(\R)$, as well as the probability measures (i.e. measurement statistics) associated with the observables.
The considerations are based on the elements of classical probability theory: the method of moments as well as the theory of weak convergence and compactness of probability measures
(see e.g. \cite{Billingsley} and \cite{BillingsleyII}).

Section \ref{generalsection}, \ref{asymptoticmeas}, and the two appendices are devoted to the above considerations.
Having investigated the general questions, we show that the moments associated with
the balanced homodyne detection fit into this scheme:
In section \ref{homodynedetector}, we give a simple expression for the observable measured by the homodyne detector as a Naimark dilation,
which we use in section \ref{moments} to determine the essential properties of the moment operators of that observable.
The purpose of section \ref{characteristic} is to demonstrate an easy (and mathematically unproblematic) way of calculating the characteristic functions for the probability measures corresponding to the coherent states, and thereby verify the similar result in \cite{Vogel} mentioned above. We do not, however, need
these characteristic functions in our moment approach.

In the final section, we gather the results together to provide a conclusion concerning the high amplitude limit.

\section{A few general notations and definitions}\label{notations}

Let $\hil$ be a complex Hilbert space and let $L(\hil)$ stand for the set of bounded operators on $\hil$. For any (not necessarily bounded) operator $A$ in $\hil$, we
denote by $D(A)$ the domain of definition of $A$. If $A$ is a selfadjoint operator, we
let $E^A:\h B(\R)\to L(\hil)$ be the unique spectral measure of $A$, where $\h B(\R)$ is the Borel
$\sigma$-algebra of $\R$. For any two vectors $\vp,\psi\in \hil$, we let $|\vp\ket\bra \psi|$ denote
the operator $\eta\mapsto \bra \psi|\eta\ket \vp$.

If $\Omega$ is a set and $\h A$ a $\sigma$-algebra of subsets of $\Omega$, a weakly (or, equivalently, strongly)
$\sigma$-additive set function $E:\h A\to L(\hil)$ with $E(B)\geq 0$ for all $B\in \h A$, and $E(\Omega)=I$, is
a \emph{semispectral measure} (also called a normalized POM). For each $\psi,\vp\in \hil$, the scalar valued set function
$\h B(\R)\ni B\mapsto \bra \psi |E(B)\vp\ket\in \C$ is then a complex measure, and we will denote it by $E_{\psi,\vp}$. For each
positive operator $T$ of trace one, we let $E_T:\h A\to [0,1]$ be defined by $E_T(B)=\tr[TE(B)]$. Then $E_T$ is a probability measure.
For a unit vector $\vp\in \hil$, we then have $E_{|\vp\ket\bra\vp|}= E_{\vp,\vp}$, and we write simply $E_\vp$ to denote this probability measure.

The operator integral \cite{Lahti} is the basic tool in our approach. If $E:\h A \to L(\hil)$ is a semispectral measure, and $f:\Omega\to \C$ a measurable function, there exists a unique (not necessarily densely defined) linear operator $L(f,E)$ on $\hil$, such that
$$
\bra \psi |L(f,E)\vp\ket = \int f \,dE_{\psi,\vp}, \ \ \psi\in \hil, \, \vp\in D(f,E),
$$
where the domain $D(f,E)$ is the set of those vectors $\vp\in \hil$ for which
the function $f$ is integrable with respect to the complex measure $E_{\psi,\vp}$ for all $\psi\in \hil$.
We will mostly use the \emph{square-integrability domain}
$$
\tilde{D}(f,E) = \{ \vp\in \hil\mid \int |f|^2\,dE_\vp<\infty\},
$$
which is a (possibly proper) subspace of $D(f,E)$ \cite{Lahti}. For convenience, we will use
the symbol $\tilde{L}(f,E)$ to denote the restriction of $L(f,E)$ to $\tilde{D}(f,E)$.

When the Hilbert space $\hil$ is associated with a
quantum system, any state of the system is represented by a positive operator $T$ of trace
one, and an observable of the system is represented by a semispectral
measure $E:\h A\to L(\hil)$ (with the measurable space $(\Omega,\h A)$ chosen suitably).
The associated probability measure $E_T$ then describes the measurement statistics
of $E$ in the state $T$. In this paper, we consider only the case $(\Omega, \h A)=(\R,\h B(\R))$.

The $k$th \emph{moment operator} of a semispectral measure $E:\h B(\R)\to L(\hil)$
is the operator integral $L(x^k,E)$ (where $x^k$ is the function $x\mapsto x^k$). Each moment
$L(x^k,E)$ is a symmetric operator.

A semispectral measure $E:\h B(\R)\to L(\hil)$ is a \emph{spectral measure}, if $E(B)$ is a projection
for all $B\in \h A$. Then all the moment operators $L(x^k,E)$ are selfadjoint and densely defined, with
$D(x^k,E)=\tilde{D}(x^k,E)$ and $L(x^{k},E)=L(x,E)^k$ for each $k\in \N$. Indeed, $E$ is then
the unique spectral measure of the selfadjoint operator $L(x,E)$.

\section{Convergence of observables in terms of their moments}\label{generalsection}

In the final section of the paper, we will see that the convergence taking place in the homodyne detection scheme
can be specified by using the moment operators of the associated semispectral measures.
The purpose of the present section is to
precisely formulate the aforementioned moment convergence in the general setting, as well as to connect it to the weak convergence
of the probability measures associated with the semispectral measures in question. Throughout the
section, $\hil$ is a complex separable Hilbert space. (Separability is needed in the main results).

\begin{definition}\label{momentlimitdef}\rm Let 
$E^n$, $n\in \N$, and $E$ be semispectral measures $\h B(\R) \to L(\hil)$. If $\h D\subset (\cap_{n,k\in \N}D(x^k,E^n))\cap D(x^k,E)$ is a dense subspace, such that
$$
\lim_{n\goto\infty} \bra \psi|L(x^k,E^n)\vp\ket = \bra \psi |L(x^k,E)\vp\ket, \ \ \ k\in \N, \, \psi\in \hil,\,\vp\in \h D
$$
(in particular, each limit in the left hand side exists), then we say that $E$ is \emph{a moment limit
for $(E^n)_{n\in \N}$ on $\h D$}. 
\end{definition}

The following observation will be useful.

\begin{proposition}\label{momentlimitprop} Let $\h D\subset \hil$ be a dense subspace, and
let $E^n$ and $E$ be semispectral measures $\h B(\R) \to L(\hil)$. Then $E$ is a moment limit for $(E^n)_{n\in \N}$ on $\h D$,
if and only if for each unit vector $\vp\in \h D$, the probability measures
$E^n_\vp$ and $E_\vp$ have all moments, and
\begin{equation}\label{probabilitymoment}
\lim_{n\goto\infty} \int x^k\, dE^n_\vp(x) = \int x^k\, dE_\vp(x), \ \ \ k\in \N, \,\vp\in \h D,\, \|\vp\|=1.
\end{equation}
\end{proposition}
\begin{proof}
If $E$ is a moment limit for $(E^n)_{n\in \N}$ on a dense subspace $\h D$, then $\h D \subset D(x^{2k},E)\subset \tilde{D}(x^k,E)$ for
any $k\in \N$, and $\h D\subset D(x^{2k},E^n)\subset \tilde{D}(x^k,E^n)$ for $n,k\in \N$, and \eqref{probabilitymoment}
clearly holds.
Assume then that there is a dense subspace $\h D$, such that each $x^k$ is $E^n_\vp$- and $E_\vp$-integrable
whenever $n\in \N$, $\vp\in \h D$, $\|\vp\|=1$, and that \eqref{probabilitymoment} holds.
Then $\h D\subset \tilde{D}(x^k,E)\subset D(x^k,E)$, and the same is true for any $E^n$ in place of $E$.
Since $\h D$ is a subspace, the condition \eqref{probabilitymoment} implies that
\begin{equation}\label{almostweak}
\lim_{n\goto\infty} \bra \psi|L(x^k,E^n)\vp\ket = \bra \psi |L(x^k,E)\vp\ket.
\end{equation}
holds for any $k\in \N$, $\vp,\psi\in \h D$ by polarization.
Let $\vp\in \h D$ and $k\in \N$. Since each $E^n$ is a semispectral measure, we have
$\|L(x^k,E^n)\vp\|^2\leq \int x^{2k} \, dE^n_\vp(x)$  (see e.g the proof of \cite[Lemma A.2]{Lahti}), and hence
\eqref{probabilitymoment} implies that the sequence $(L(x^k,E^n)\vp)_{n\in\N}$ is bounded.
It follows that \eqref{almostweak} holds for all $k\in \N$, $\psi\in \hil$, $\vp\in \h D$ (see e.g \cite[Theorem 2, p. 47]{Akhiezer}),
so that $E$ is a moment limit for $(E^n)_{n\in \N}$ (on $\h D$).
\end{proof}

\begin{remark}\rm A sequence $(E^n)$ of semispectral measures
can have various moment limits $E$ on a subspace $\h D$. For a trivial example, take a constant sequence $E^n=\mu I$,
with $\mu$ a probability measure which is not determined by its moments, i.e. there exists another probability measure with the same moments.
\end{remark}

As the preceding remark demonstrates, the uniqueness of the moment limit of a sequence of
observables is connected to the limiting probability measures being determined
by their moment sequences.
Recall that 
a positive measure $\mu:\h B(\R)\to [0,\infty)$ is called \emph{determinate}, if each moment $M_k:=\int x^k\,d\mu(x)$, $k=0,1,2\ldots$,
 exists (and is finite) and the moment sequence $(M_k)$ determines the measure $\mu$,
i.e there is no other measure $\nu:\h B(\R)\to [0,\infty)$ such that $\int x^k \, d\nu(x) = M_k$ for all $k=0,1,2,\ldots$.

\begin{definition}\rm  Let $E:\h B(\R)\to L(\hil)$ be a semispectral measure, and
let $\h D\subset \hil$ be a set. If the positive measure $E_\vp$ is determinate for each $\vp\in \h D$, then
$E$ is \emph{$\h D$-determinate}.
\end{definition}

\begin{remark}\label{determinacy}\rm
It is easy to see that for a dense subspace $\h D$, a $\h D$-determinate semispectral measure $E$ is \emph{determinate},
i.e. the only semispectral measure having the same moment operators as $E$ is $E$ itself. In fact,
assume that $E$ is $\h D$-determinate with $\h D$ a dense subspace, and let $E':\h B(\R)\to L(\hil)$ be a semispectral measure such that
$L(x^k,E)= L(x^k,E')$ for all $k\in \N$. Since $E_\vp$ is determinate
for each $\vp\in \h D$, we have, in particular, that $\int x^k\,dE_\vp(x)<\infty$ for all $\vp\in \h D$. Hence,
$\h D\subset \tilde{D}(x^k,E) \subset D(x^k,E)=D(x^k,E')$ for all $k\in \N$. Now $E_\vp=E'_\vp$ for all $\vp\in \h D$, since
$E_\vp$ is determinate and $L(x^k,E)\vp=L(x^k,E')\vp$ for all $\vp\in \h D$. Using polarization and the density of $\h D$,
we get $E=E'$, proving that $E$ is determinate.

If $\h D$ is not dense, a $\h D$-determinate
semispectral measure need not be determinate. For an example, let $P\in L(\hil)$ be a projection, $\mu$ a determinate probability
measure on $\R$ (thus having all moments), and define $E^\nu:\h B(\R)\to L(\hil)$ by
$E^\nu(B) = \mu(B)P+\nu(B)(I-P)$ for each probability measure $\nu$ with $\int |x| d\nu(x)=\infty$.
Now each such $E^\nu$ is a $\h D$-determinate semispectral measure, but $D(x^k,E^\nu)=\tilde{D}(x^k,E^\nu) =P(\hil)$,
and $L(x^k,E^\nu)$ is the operator $P(\hil)\ni \vp\mapsto (\int x^k d\mu(x))\vp\in \hil$, so
that the moment sequence of $E^\nu$ does not depend on $\nu$.
\end{remark}

\begin{remark} \label{determinacy2}\rm
Although the concept of $\h D$-determinate semispectral measure may seem somewhat artificial
compared to
the natural definition of determinate semispectral measure, many existing physically relevant examples of the latter actually fall also into the former category.
In particular, it was proved in \cite{DvurecenskijI} that the polar margins of
a certain physically relevant phase space semispectral measures are determinate, but
from the proof we see that these semispectral measures are actually $\h D$-determinate, where $\h D\subset L^2(\R)$ is the
linear span of the Hermite functions. In \cite{DvurecenskijII} it was shown that the cartesian margins of the
same phase space semispectral measures are $C^\infty_0(\R)$-determinate,
where $C^\infty_0(\R)$ is the space of complex valued compactly supported infinitely differentiable functions.
\end{remark} 

The following result emphasizes the relevance of the concept of a $\h D$-determinate semispectral measure in our considerations.

\begin{proposition}\label{uniquenessofmomentlimit}
Let $E^n$, $n\in \N$, and $E$ be semispectral measures $\h B(\R)\to L(\hil)$, and $\h D\subset \hil$ be
a dense subspace, such that $E$ is a moment limit for $(E^n)_{n\in \N}$ on $\h D$. If $E$ is $\h D$-determinate, then $E$ is the only moment limit for  $(E^n)_{n\in \N}$ on $\h D$.
\end{proposition}
\begin{proof}
According to proposition \ref{momentlimitprop}, any other moment limit $E'$ on $\h D$ satisfies
$\int x^k\, dE_\vp=\int x^k \, dE'_\vp$ for all $\vp \in \h D$. Since $E$ is $\h D$-determinate,
it follows that $E_\vp = E'_\vp$ for all $\vp\in \h D$, which implies $E=E'$ by polarization and
the density of $\h D$.
\end{proof}

It is often difficult to determine, whether a given semispectral measure is $\h D$-determinate for a given
subspace $\h D$. In some applications, however, the relevant positive measures possess the following stronger
property \cite{DvurecenskijI}: A measure $\mu:\h B(\R)\to [0,\infty)$ is \emph{exponentially bounded}, if
$\int e^{a|x|} \,d\mu(x)<\infty$ for some $a>0$. As mentioned in \cite{DvurecenskijI}, such a measure is determinate
by e.g. \cite[Theorem II.5.2]{Freud}.

For a semispectral measure $E:\h B(\R)\to L(\hil)$, we let $\h E_E$ denote the set of vectors $\vp\in \hil$ for which $E_\vp$ is
exponentially bounded. Later we will need the following observation.

\begin{lemma}\label{exponentialboundednesslemma} For any semispectral measure $E:\h B(\R)\to L(\hil)$,
the set $\h E_E$ is a subspace of $\hil$, and $E$ is $\h E_E$-determinate.
\end{lemma}
\begin{proof}
If $\vp,\psi\in \h E_E$, and $c_1,c_2\in \C$, we have
\beat
E_{c_1 \vp+c_2\psi}(B) &=& \|E(B)^{\frac 12}(c_1\vp+c_2\psi)\|^2
\leq  (|c_1|\|E(B)^{\frac 12}\vp\|+|c_2|\|E(B)^{\frac 12}\psi\|)^2\\
&\leq&  2|c_1|^2E_{\vp}(B)+2|c_2|^2E_{\psi}(B)
\eeqat
for all $B\in \h B(\R)$. This clearly implies that also $E_{c_1\vp+c_2\psi}$ is exponentially bounded, showing
that $\h E_E$ is a subspace. The last statement follows from the previously mentioned fact that an exponentially
bounded measure is determinate.
\end{proof}

Having investigated the moment convergence in detail, we now want to describe this
convergence at the level of the probability distributions themselves.
The motivation for this comes partly from the fact that Vogel \cite{Vogel} claims that the relevant convergence concept in the homodyne detector problem is the weak convergence
of all the probability measures associated with the measured observables. In the definition below, we formulate this convergence
in terms of the observables themselves. This formulation has been used
e.g. in \cite[Theorem 1]{Blum} (where it appeared without any specific name or physical context).
One should recall the definition of the weak convergence of probability measures \cite[p. 11]{Billingsley}:
a sequence $(\mu_n)$ of probability measures on $\h B(\R)$ converges weakly to a probability measure
$\mu:\h B(\R)\to [0,1]$ if $\lim_{n\goto\infty} \int f \,d\mu_n =\int f\,d\mu$ for all bounded continuous functions $f:\R\to \R$.

For each $B\subset \R$, we let $\partial B$ denote the
boundary of $B$, i.e. $\partial B=\cc B \cap \cc{\R\setminus B}$. 
This notation is needed in a characterization of the weak convergence,
which says that a sequence $(\mu_n)$ of probability measures $\h B(\R)\to [0,1]$ converges weakly to a
probability measure $\mu:\h B(\R)\to [0,1]$, if and only if $\lim_{n}\mu_n(B)=\mu(B)$ for all
$B\in \h B(\R)$ with $\mu(\partial B)=0$ (see e.g. \cite[Theorem 2.1]{Billingsley}).
One should also recall that 
a sequence of probability measures
converges weakly, if and only if the associated sequence of distribution functions converges pointwise at all points
where the limiting distribution function is continuous \cite[p. 335-336]{Billingsley}.

\begin{definition}\label{weakdef}\rm Let $\hil$ be a Hilbert space, and let $E^n:\h B(\R)\to L(\hil)$ be a semispectral measure for each
$n\in \N$. We say that the sequence $(E^n)$ converges to a semispectral measure $E:\h B(\R)\to L(\hil)$
\emph{weakly in the sense of probabilities}, if
$$ \lim_{n\goto\infty} E^n(B) = E(B)$$ in the weak operator topology, for all $B\in \h B(\R)$ such that $E(\partial B)=0$.
\end{definition}

The following proposition characterizes this convergence. The proof of this
result is given in Appendix A (in the more general context of a metric space).

\begin{proposition}\label{weakconvergence} Let
$E^n,E:\h B(\R)\to L(\hil)$, $n\in \N$, be semispectral measures. Then the following conditions are equivalent.
\begin{itemize}
\item[(i)] $(E^n)$ converges to $E$ weakly in the sense of probabilities;
\item[(ii)] for each positive operator $T$ of trace one, the sequence $(E^n_T)$ of probability measures converges
weakly to $E_T$;
\item[(iii)] there exists a dense subspace $\h D\subset \hil$, such that the sequence $(E^n_\vp)$ of
probability measures converges weakly to $E_\vp$ for any unit vector $\vp\in \h D$;
\item[(iv)] $\lim_{n\goesto \infty} L(f,E^n) = L(f,E)$ in the weak operator topology for each bounded continuous
function $f:\R\to \R$.
\end{itemize}
\end{proposition}

The main result of this section, Proposition \ref{momentasymptotic} below, gives a connection between the moment limit and
convergence in the sense of probabilities, which is analogous to the corresponding result for probability measures
\cite[Theorem, p. 540]{Frechet}.
The proof of part (a) of the proposition is based on the following result, which uses the concept of relative
compactness as it appears in probability theory:
a family $\h P$ of probability measures $\nu: \h B(\R)\to [0,1]$ is called
\emph{relatively compact}, if every sequence of elements of $\h P$ contains a weakly convergent subsequence
(see \cite[p. 35]{Billingsley}). The proof of Proposition \ref{weaklycompact2} is given in Appendix B.

\begin{proposition}\label{weaklycompact2} Let $\h D\subset\hil$ be a dense subspace,
and let $\h M$ be a collection of semispectral measures $E:\h B(\R)\to L(\hil)$.
Suppose that the set $\{E_{\vp}\mid E\in \h M\}$ of probability measures is relatively compact for each unit vector
$\vp\in \h D$.
Then every sequence of elements of $\h M$ contains a subsequence which converges weakly in the sense of probabilities.
\end{proposition}

\begin{proposition}\label{momentasymptotic}
Let $E^n:\h B(\R)\to L(\hil)$ be a semispectral measure for each $n\in \N$. Assume that there is a dense
subspace $\h D\subset \cap_{n,m\in \N} D(x^m,E^n)$, such that the limit
$$
\lim_{n\goto\infty} \bra \vp|L(x^m,E^n)\vp\ket
$$
exists in $\R$ for each $m\in \N$ and $\vp\in \h D$.
\begin{itemize}
\item[(a)] There exists a semispectral measure $E:\h B(\R)\to L(\hil)$, which is a moment limit for $(E^n)_{n\in \N}$ on $\h D$.
\item[(b)] Suppose that $E$ is $\h D$-determinate. Then $E$ is the only moment limit for $(E^n)_{n\in\N}$ on $\h D$, and
the sequence $(E^n)_{n\in \N}$ converges to $E$ weakly in the sense of probabilities. 
\end{itemize}
\end{proposition}
\begin{proof}
By assumption, $$\sup_{n\in \N} \int x^2 \, dE^n_{\vp}(x) <\infty$$ for all $\vp\in \h D$. This implies
that $\{ E^n_\vp \mid n\in \N\}$ is relatively compact for each unit vector $\vp\in \h D$.
(See e.g. the first paragraph in the proof of \cite[Theorem 30.2, p. 408]{Billingsley}, and
Prohorov's theorem \cite[Theorem 6.1, p. 37]{Billingsley}.)
Hence, we can apply Proposition \ref{weaklycompact2} to get a subsequence $(E^{n_k})_{k\in \N}$ converging to a
semispectral measure $E:\h B(\R)\to L(\hil)$ weakly in the sense of probabilities.
Fix a unit vector $\vp\in \h D$. By Proposition \ref{weakconvergence},
the sequence $(E^{n_k}_\vp)_{k\in \N}$ of probability measures converges to $E_\vp$ weakly.
Since, in addition,
$$\lambda_m(\vp) :=\lim_{n\goto\infty} \int x^m\, dE^{n}_\vp(x)=\lim_{k\goto\infty} \int x^m\,dE^{n_k}_\vp(x)$$ exists for all
$m\in \N$ by assumption, it follows from the proof of Theorem 30.2 of \cite[p. 408]{BillingsleyII} (or directly
from \cite[Theorem 4 A]{Rao})
that $\lambda_m(\vp) = \int x^m\, dE_\vp(x)$. But by Proposition \ref{momentlimitprop} this means that $E$ is a moment limit
for $(E^n)_{n\in \N}$ on $\h D$. Hence, (a) is proved.

Assume then that $E$ is $\h D$-determinate. It follows from Proposition \ref{uniquenessofmomentlimit} 
that $E$ is the only moment limit for $(E^n)_{n\in \N}$ on $\h D$. Let $\vp\in \h D$ be a unit vector. Then the numbers
$\int x^m\,dE_\vp(x)$ are the moments for only one probability measure (which is $E_\vp$), so it follows directly from
\cite[Theorem  30.2]{BillingsleyII} (or \cite[Theorem 6]{Rao})
that $(E^n_\vp)_{n\in \N}$ converges to $E_\vp$ weakly. By proposition \ref{weakconvergence}, this entails that
$(E^n)_{n\in \N}$ converges to $E$ weakly in the sense of probabilities.
\end{proof}

\section{An ''asymptotic measurement'' scheme}\label{asymptoticmeas}

Suppose that we have a measurement setup which can be configured to various modes $\h M_n$, $n=0,1,2,\ldots$,
each of which is supposed to constitute a measurement, which is somehow an approximation of the observable we actually
want to measure, the approximation becoming more and more accurate as $n$ increases. This is, of course, somewhat vague,
but it can be given a reasonably precise meaning by considering the moments of the measurement outcome statistics. In the following,
we number the steps so that they can be referred to later.

\begin{enumerate}
\item\label{item1}
According to quantum measurement theory, each measurement $\h M_n$ can (in principle) be associated with a semispectral
measure $E^n$, acting on
a (common) Hilbert space $\hil$. Suppose that we have a fixed set of measurement preparations to be used for the calibration
of the measurement setup (i.e. determination of the measured observable), and assume that this set can be identified
with the unit ball $\h D_1$ of some dense subspace $\h D\subset \hil$.
\item\label{item2}
Then each measurement $\h M_n$, performed with an input state $\vp\in \h D_1$, produces the probability distribution $E^n_\vp$,
whose moments $M_{n,\vp}^k$, $k\in \N$, can be calculated from the measurement data. (Notice that
this can be done without needing to know the actual form of the semispectral measure $E^n$). After this point, the process
obviously works only if we can convince ourselves that all the moments $M_{n,\vp}^k$ exist.
\item\label{item3}
Assume then that, once a sufficient number of the measurements $\h M_n$ have been performed, one arrives at the conclusion that
$$
\lim_{n\goesto \infty} M_{n,\vp}^k \text{ exists for each } k\in \N,\, \vp\in \h D_1.
$$
Then Proposition \ref{momentasymptotic} (a) says that there exists at least one observable $E$, such that
the numbers obtained from the limits $\lim_{n\goesto \infty} M_{n,\vp}^k$ coincide with the corresponding moments
of the measurement statistics of $E$. Thus, in the level of moments, this scheme describes a valid measurement.
\item\label{item4}
However, this approach cannot lead to a measurement of a unique quantum observable, unless
the moments obtained as the above mentioned limits actually determine uniquely the probability distributions
from which they arise.
Hence, it is natural to require that each measure $E_\vp$, $\vp\in \h D_1$ be determinate, i.e. $E$ be $\h D$-determinate, in which
case Proposition \ref{momentasymptotic} (b) implies that exactly one observable $E$ can be associated
with the limit statistics corresponding to the set $\h D_1$ of calibration states.
\item\label{item5}
One additional condition could be imposed in order to make the above reasoning more solid. Namely, also the semispectral measures
$E^n$ could be required to be $\h D$-determinate. In view of Proposition \ref{momentasymptotic}, this is not necessary, but
it would nevertheless ensure that for each $n$, the moments $M_{n,\vp}^k$ used in
the above process completely determine the measurement statistics of $\h M_n$.
\end{enumerate}

Thus, if all the above conditions are satisfied, then the scheme can be considered as a measurement of
the unique moment limit $E$ of the sequence $(E^n)$ on $\h D$. Moreover, it is important to note that then $(E^n)_{n\in \N}$
converges to $E$ weakly in the sense of probabilities (Proposition \ref{momentasymptotic} (b)), which means
(see Proposition \ref{weakconvergence}) that the sequence $(E^n_T)$ of outcome probability measures converges
weakly to $E_T$ for \emph{any} input state $T$ (positive operator of unit trace), not just for the states used in the calibration.

As we will see in the following sections, all the above requirements are satisfied for a simple theoretical model of the
balanced homodyne detector, when the linear span of the set of coherent states are used in the calibration.

\begin{remark}\rm
The requirement that the observables be $\h D$-determinate is quite restrictive, and may be difficult
to verify without knowing the explicit form for the observables. In some cases, the question
might be resolved by using a criterion from the extensive literature concerning the classical moment problem (see for instance
\cite{AkhiezerII} or \cite{Freud}). For example, if all the limiting moments $M_{k,\vp} := \lim_{n\goesto \infty} M_{n,\vp}^k$
satisfy the condition $\liminf_{k\goto\infty} \frac {1}{2k} M_{2k, \vp}^{\frac 1 {2k}}<\infty$, then the
limiting observable $E$ is $\h D$-determinate
(see e.g. \cite[Theorem II.5.1]{Freud}). In our example of the homodyne detection, this condition actually holds, but we will
not need to state it explicitly, as the relevant measures have the stronger property of exponential boundedness (see the
proof of Lemma \ref{quadraturedeterminacy}). The fact that exponential boundedness implies the above moment
condition can be seen e.g. from the proof of \cite[Theorem II.5.2]{Freud}.
\end{remark}

\section{The balanced homodyne detector}\label{homodynedetector}

Now we proceed to the description of the balanced homodyne detection scheme.
A simple description of this detector can be given as follows: a (single mode) signal field with the Hilbert space
$\hil$ is coupled with an auxiliary field (with the Hilbert space $\hil_{aux}$) via a 50-50-beam splitter
described by a certain unitary operator $U$.
The auxiliary field is initialized in the coherent state $|z\ket$. The photon number difference for the two output
ports of the beam splitter, divided by the strength $|z|$ of the auxiliary field, is then detected. (We consider only
ideal photodetectors in our scheme.)

Assuming that $\hil$ and $\hil_{aux}$ are separable complex Hilbert spaces, we use the following usual notations.
We fix orthonormal bases of the form $\{|n\ket \mid n\in \N\}$ for both $\hil$ and $\hil_{aux}$, representing the photon
number states. We denote
\bet
|nm\ket = |n,m\ket :=|n\ket\otimes |m\ket\in \hil\otimes \hil_{aux}, \ \ n,m\in \N.
\eeqt
Let $a,a^*$ and $b,b^*$ be the creation and annihilation operators for
the aforementioned bases of $\hil$ and $\hil_{aux}$, respectively, and let $N=a^*a$ and $N_{aux}=b^*b$,
be the photon number operators for the two modes. The operators $a,a^*,b,b^*,N,N_{aux}$ are considered as being defined
on their natural domains, e.g.
\beat
D(a) = D(a^*) &=& \{ \vp\in \hil \mid \sum_{n\in \N} n|\bra n|\vp\ket|^2<\infty\};\\
D(N) = D(a^*a) &=& \{ \vp\in \hil \mid \sum_{n\in \N} n^2 |\bra n|\vp\ket|^2<\infty\}.
\eeqat

For any $z\in \C$ the \emph{coherent state} $|z\ket\in \hil$ is defined by
\bet
|z\ket = e^{-\frac 12 |z|^2}\sum_{n=0}^\infty\frac{z^n}{\sqrt{n!}} |n\ket,
\eeqt
and we use the same symbols for the coherent states in $\hil_{aux}$. Also, we use the shorthand
$|z,w\ket = |z\ket \otimes |w\ket\in \hil\otimes \hil_{aux}$.
The subspace $D_{coh} := \lin \{|z\ket \mid z\in \C\}$ is dense in $\hil$, and so is the corresponding subspace
$D_{coh}^{aux}$ in $\hil_{aux}$. (Here the symbol ''lin'' denotes the (algebraic) linear span of the set in question.)
The (algebraic) tensor product $\Dcoh:=D_{coh}\otimes D_{coh}^{aux}$ can
be identified with ${\rm lin} \{|\beta,z\ket \mid \beta,z\in \C\}$, which is dense in $\hil\otimes\hil_{aux}$.
The following standard formulas hold:
\beat
e^{it N}|z\ket &=& |e^{it}z\ket, \ \ e^{itN_{aux}}|z\ket = |e^{it}z\ket,\\
\bra z|z'\ket &=& e^{-\frac 12 (|z|^2+|z'|^2)+\cc z z'}.
\eeqat

Denote by $Q$ and $P$ the signal field quadrature operators
$\frac{1}{\sqrt{2}}\cc{(a^*+a)}$ and $\frac{i}{\sqrt{2}}\cc{(a^*-a)}$, respectively. Here the bar stands for the
closure of an operator, so that e.g. $Q$ is the unique selfadjoint extension of the essentially selfadjoint
symmetric operator $\frac 1 {\sqrt 2} (a^*+a)$ (See \cite[Chapter IV]{Putnam} or \cite[Chapter 12]{Birman}
for details concerning the domains of these very extensively studied operators.)
The set of states for the signal field is denoted by $\h S(\hil)$, i.e. $\h S(\hil)$ consists of those $T\in L(\hil)$,
which are positive and of trace one.

For the signal field, define the \emph{rotated quadrature operators} $Q_\theta$, with $\theta\in [0,2\pi)$, via
\be\label{rotatedquadrature}
Q_\theta = e^{i\theta N}Qe^{-i\theta N} = \frac 1 {\sqrt 2} (\cc{e^{-i\theta} a+e^{i\theta} a^*})
\eeq
In particular, each $Q_\theta$ is selfadjoint on its domain
$D(Q_\theta)= e^{i\theta N}D(Q)\supset D(a)=D(a^*)$, and the restriction
$Q_\theta|_{D(a)} = \frac 1 {\sqrt 2} (e^{-i\theta} a +e^{i\theta} a^*)$ is essentially selfadjoint.
The ordinary quadratures are given by $Q_{0} = Q$ and $Q_{\frac \pi 2}=P$. We will need the following fact.

\begin{lemma}\label{quadraturedeterminacy}
For all $\theta\in [0,2\pi)$, the spectral measure of $Q_\theta$ is $D_{coh}$-determinate.
\end{lemma}
\begin{proof} Let $\theta\in [0,2\pi)$. By the definition \eqref{rotatedquadrature} of $Q_\theta$, the
spectral measure $E^{Q_\theta}$ is $X\mapsto e^{i\theta N}E^Q(X)e^{-i\theta N}$.
Let $\beta\in \C$, and consider the probability measure $E^{Q_\theta}_{|\beta\ket}$, which has the simple form
$E^{Q_\theta}(X) = \bra \beta'|E^Q(X)|\beta'\ket$, where $|\beta'\ket = |e^{-i\theta}\beta\ket$. In the coordinate
representation ($\hil \simeq L^2(\R)$, $|n\ket\mapsto h_n$, with $h_n$ the $n$th Hermite function), $Q$ is
the multiplication by the function $x\mapsto x$, and $|\beta'\ket$ is the vector $W(q,p)h_0$, where $\beta'=\frac 1 {\sqrt 2}(q+ip)$,
and $W(q,p)$ is the Weyl operator, so that $(W(q,p)h_0)(x) = \pi^{-\frac 14} e^{-i\frac 12 qp+ipx}e^{-\frac 12(x-q)^2}$.
Hence, $E^{Q_\theta}_{|\beta\ket}(X) = \frac 1 {\sqrt \pi}\int_X e^{-(x-q)^2} dx$, which immediately implies
that $\int e^{a|x|}\, dE^{Q_\theta}_{|z\ket} <\infty$ for any $a>0$. Hence, $|\beta\ket\in \h E_{E^{Q_\theta}}$ for each
$\beta\in \C$. It then follows from Lemma \ref{exponentialboundednesslemma} that $E^{Q_\theta}$ is $D_{coh}$-determinate.
\end{proof}

In quantum optics, it is typically asserted that a (lossless) 50-50 beam splitter with phase shift $\phi$
can be described by a unitary operator $U_{\phi}$, satisfying
\be\label{UcohI}
U_{\phi}|\beta,z\ket=|\tfrac 1 {\sqrt 2}(\beta-e^{i\phi}z), \tfrac 1 {\sqrt 2} (e^{-i\phi}\beta+ z\ket).
\eeq
In particular, $U (D_{coh}\otimes D_{coh}^{aux})=D_{coh}\otimes D_{coh}^{aux}$. In terms of annihilation operators, \eqref{UcohI} implies
\beat
U_\phi^*(a\otimes I)|_{\Dcoh}U_\phi &=& \frac 1 {\sqrt 2} (a-e^{i\phi}b)|_{\Dcoh};\\
U_\phi^*(I\otimes b)|_{\Dcoh}U_\phi &=& \frac 1 {\sqrt 2} (e^{-i\phi} a+b)|_{\Dcoh}.
\eeqat

Using these equations, we see that the restriction
of the operator $I\otimes N_{aux}-N\otimes I_{aux}$ to $\Dcoh$ is unitarily equivalent to
the corresponding restriction of $(e^{-i\phi}a\otimes b^*+e^{i\phi}a^*\otimes b)$, the equivalence being carried by $U$. Now
both of these operators are essentially selfadjoint, and the closure of the 
former is the photon number difference observable, which we denote by $N_-$. Hence,
$$
\frac {1}{\sqrt{2}} U_\phi^*N_-U_\phi = A_\phi,
$$
where $A_\phi$ is the closure of the operator $\frac {1}{\sqrt{2}}(e^{-i\phi}a\otimes b^*+e^{i\phi} a^*\otimes b)$.
Furthermore, $A_\phi = I \otimes e^{-i\phi N_{aux}}\,A\, I\otimes e^{i\phi N_{aux}}$, where $A:=A_0$.

According to the scheme, the detection observable for the two-mode field is the photon difference $N_{-}$, divided by the amplitude $|z|$
of the input $|z\ket$ of the auxiliary field. For later convenience, we scale it with the factor $\frac 1 {\sqrt 2}$, so
that the detection is represented by the spectral measure 
\bet
\hB(\R)\ni B\mapsto E^{(\sqrt 2 |z|)^{-1}N_{-}}(B)\in L(\hil\otimes \hil_{aux})
\eeqt
of the selfadjoint operator $(\sqrt 2 |z|)^{-1}N_{-}$.
If $T$ is the input state of the signal field, the two-mode field state after the beam splitter is $U_\phi T\otimes |z\ket \bra z|U_\phi^*$.
Hence, the detection statistics are given by the probability measures
\bet
B\mapsto p_T^{z,\phi}(B):=\tr[T\otimes |z\ket\bra z| U^*E^{(\sqrt 2 |z|)^{-1}N_{-}}(B)U], \ \ T\in \h S (\hil), \, z\in \C.
\eeqt
Because of the unitary equivalences $\frac 1 {\sqrt 2} U_\phi^*N_{-}U_\phi =A_\phi= I \otimes e^{-i\phi N_{aux}}\,A\, I\otimes e^{i\phi N_{aux}}$,
we get
\be\label{detectionstat}
p_T^{z,\phi}(B)= \tr[T\otimes |z\ket\bra z| E^{|z|^{-1} A_\phi}(B)] = \tr [T\otimes |e^{i\phi}z\ket\bra e^{i\phi}z| E^{|z|^{-1}A}(B)],
\eeq
reflecting the fact that the phase shift in the beam splitter can be realized by shifting the phase of the
auxiliary field. Accordingly, we put $\phi=0$ in the subsequent discussions, and write instead $z=re^{i\theta}$, so that $\theta$ represents the phase difference between the two input modes. The beam splitter is now $U:=U_0$, with
\be\label{Ucoh}
U|\beta,z\ket=|\tfrac 1 {\sqrt 2}(\beta-z), \tfrac 1 {\sqrt 2} (\beta+ z)\ket.
\eeq

Since the signal field is considered as the input for the actual homodyne detector with the fixed auxiliary state $|z\ket=|re^{i\theta}\ket$,
the observable $E^z:\h B(\R)\to L(\hil)$ being measured by the detector is now uniquely determined, as a semispectral measure, by the relation
\be\label{theobservable}
\tr[TE^z(B)] = \tr[T\otimes |z\ket\bra z| E^{|z|^{-1} A}(B)], \ \ B\in \h B(\R), \, T\in \h S(\hil).
\eeq
(The reader may wish to consult \cite{OQP} for background information on quantum measurement schemes and semispectral measures; the
balanced homodyne detection is briefly discussed on pp. 194-195.)

Let $V_z:\hil\to \hil\otimes \hil_{aux}$ be the linear isometry $\vp\mapsto \vp\otimes |z\ket$.
In the formulation of measurement dilations, the relation \eqref{theobservable} says that
$(\hil\otimes \hil_{aux}, E^{|z|^{-1} A}, V_z)$ is a Naimark dilation of the semispectral measure $E^z$, i.e.
\be\label{dilation}
E^z(B) = V_z^*E^{|z|^{-1} A}(B)V_z, \ \ B\in \h B(\R).
\eeq
(For a brief exposition of measurement dilations, see \cite[Section 2]{LahtiIV}.)

The idea in the homodyne detector is that in the high amplitude limit, i.e. the limit $|z|\goto\infty$,
the measurement statistics of the observables $E^z$ begins to resemble the corresponding statistics of the
rotated quadrature $Q_\theta$, where $\theta$ is the fixed phase of the complex number $z$. As announced above, we
approach this problem by considering the moments of these semispectral measures.

\section{The moment operators of $E^z$}\label{moments}

In this section, we determine the relevant properties of the moment operators $L(x^k, E^z)$. Most of the
discussion concerning the relevance of the results is postponed to the concluding section \ref{conclusion}.
For simplicity, we let $\tilde{L}(x^k,E^z)$ denote the restriction of $L(x^k,E^z)$ to $\tilde{D}(x^k,E^z)$.

\begin{proposition}\label{sqrprop} Let $z\in \C$ and $k\in \N$.
\begin{itemize}
\item[(a)] $\tilde{L}(x^k,E^z)=|z|^{-k}V_z^*A^kV_z$.
\item[(b)] $D(a^k)\subset \tilde{D}(x^k,E^z)$. In particular, $\tilde{L}(x^k,E^z)$ is densely defined.
\end{itemize}
\end{proposition}
\begin{proof} Since $E^{|z|^{-1}A}$ is a spectral measure, we have $L(x^k,E^{|z|^{-1}A}) = |z|^{-k}A^k$, so
(a) follows immediately from Theorem in \cite[Section III A]{LahtiIII}. If $\vp\in D(a^k)$, then
\bet
V_z\vp=\vp\otimes |z\ket\in D(a^k)\otimes D(b^k)\subset D((A|_{D(a)\otimes D(b)})^k)\subset D(A^k)=V_z\tilde{D}(x^k,E^z)
\eeqt
by (a), so $\vp\in \tilde{D}(x^k,E^z)$. This proves (b).
\end{proof}

The following result gives the first two moments explicitly.

\begin{proposition}\label{fstsndmoments} Let $z=r e^{i\theta}$, with $r>0$, $\theta\in [0,2\pi)$.
\begin{itemize}
\item[(a)] $\tilde{L}(x,E^{z})\supset Q_\theta|_{D(a)}$.
\item[(b)] $\tilde{L}(x^2,E^{z})\supset (Q_\theta|_{D(a)})^2+\tfrac 12 r^{-2} N$.
\end{itemize}
\end{proposition}
\begin{proof} Let $\vp\in D(a)$, so that $\vp\in \tilde{D}(x,E^z)$ by Proposition \ref{sqrprop} (b). Let $\psi\in \hil$. Part (a) of
the same proposition now gives
\bet
\bra \psi |\tilde{L}(x,E^z)\vp\ket = \tfrac 1 {\sqrt 2} |z|^{-1}\bra \psi \otimes |z\ket |(a\otimes b^*+a^*\otimes b)\vp\otimes |z\ket\ket
= \bra\psi|\tfrac 1 {\sqrt 2} (e^{-i\theta}a+e^{i\theta} a^*)\vp\ket.
\eeqt
This proves (a). To prove (b), note first that since $D(N)=D(a^2)=D(a^*a) = D(aa^*) = D((a^*)^2)$, we have
$D(N)\subset D((Q_\theta|_{D(a)})^2)$, and hence the domain of the operator 
$(Q_\theta|_{D(a)})^2+\frac 12 r^{-2} N$ is $D(N)$.
Now let $\vp\in D(a^2)$ and $\psi\in \hil$. Using Proposition
\ref{sqrprop} again (and noticing the fact that $(A|_{D(a)\otimes D(b)})^2\subset A^2$), we get
\beat
\bra \psi |\tilde{L}(x^2,E^z)\vp\ket &=& \tfrac 12 |z|^{-2}\bra \psi \otimes |z\ket |(a\otimes b^*+a^*\otimes b)^2\vp\otimes |z\ket\ket\\
&=& \tfrac 12 |z|^{-2}\bra \psi | (\cc z^2a^2+|z|^2aa^*+(|z|^2+1)a^*a+z^2(a^*)^2)\vp\ket\\
&=& \tfrac 12 \bra \psi | (e^{-2i\theta}a^2+e^{2i\theta}(a^*)^2+aa^*+a^*a+r^{-2}a^*a)\vp\ket\\
& =& \bra \psi |([\tfrac 1 {\sqrt 2} (e^{-i\theta}a+e^{i\theta}a^*)]^2+\tfrac 12 r^{-2} a^*a)\vp\ket.
\eeqat

\end{proof}

\begin{corollary} The \emph{intrinsic noise operator} $\h N(E^z):=L(x^2,E^z)-L(x,E^z)^2$ of the observable
$E^z$ is the selfadjoint operator $\frac 12 \frac 1 {|z|^2} N$. In particular,
$D(L(x^2,E^z))\cap D(L(x,E^z)^2) = D(a^2)$.
\end{corollary}
\begin{proof}
Since $D(N)\subset D((Q_\theta|_{D(a)})^2)$, we have $D(N)\subset D(L(x,E^z)^2)$ by part (a) of the preceding proposition.
In addition, $D(a^2)\subset D(x^2,E^z)$ by part (b) of that proposition. Hence,
$D(N) \subset D(L(x^2,E^z)-L(x,E^z)^2)$, so that $\frac 12 \frac 1 {|z|^2} N\subset L(x^2,E^z)-L(x,E^z)^2$
by the proposition. But $L(x^2,E^z)-L(x,E^z)^2$ is symmetric (even positive), and $\frac 12 \frac 1 {|z|^2} N$
is selfadjoint, so that $\tfrac 12 \frac 1 {|z|^2} N= L(x^2,E^z)-L(x,E^z)^2$.
\end{proof}

\begin{remark}\rm Proposition \ref{fstsndmoments} (a) reflects the fact that no $E^z$ is a spectral
measure (for any spectral measure is well known to have zero intrinsic noise). A perhaps somewhat less well-known
fact is that a semispectral measure with zero intrinsic noise, and a selfadjoint first moment operator, is
necessarily a spectral measure, and thereby completely determined, as a semispectral measure,
by its first and second moment operators (see \cite[Theorem 5]{Kiukas}).
By the above corollary, we have $\h N(E^z)\vp\goesto 0$
as $|z|\goesto 0$ for any $\vp\in D(\h N(E^z))=D(N)$, so one might expect that the first two moments
of the detection statistics indeed determine the observable measured in the high amplitude limit.
However, due to the problems concerning the domains of the moment operators, we were not able to
make this reasoning rigorous, and therefore chose instead to consider entire moment sequences
in the ''weak sense'', as will be shown in the last section.
\end{remark}

\begin{remark} \rm In view of the tomographic application, there is also another
aspect in the above mentioned fact that it is not enough to
consider only the first moment operator. Namely, one cannot reconstruct a state
by knowing only the averages of all the quadratures in that state. As a simple example, consider any
two different vector states $|\vp\ket\bra \vp|$, $|\psi\ket\bra\psi|$, for which $\vp,\psi\in D(Q)\cap D(P)$,
$\bra \vp|Q\vp\ket = \bra \psi|Q\psi\ket$, and $\bra \vp|P\vp\ket = \bra \psi|P\psi\ket$. [For example, one
can take, in the coordinate representation, $\vp(x) = f(x)e^{i\phi(x)}$,
$\psi(x) = f(x)e^{-i\phi(-x)}$, where $f$ is a positive even function, and $\phi:\R\to [0,2\pi)$ such
that $x\mapsto \theta(x) +\theta(-x)$ is not a constant mod $2\pi$.]
Since $D(Q)\cap D(P)=D(a)$, it follows that
$\bra \vp|a\vp\ket = \bra \psi|a\psi\ket$ and $\bra \vp|a^*\vp\ket = \bra \psi|a^*\psi\ket$, and, consequently
$\bra \vp|Q_\theta\vp\ket = \bra \psi|Q_\theta\psi\ket$ for all $\theta\in [0,2\pi)$. Hence,
the averages do not distinguish between these states.
\end{remark}


For the general moment operator $L(x^k,E^z)$, the following result is sufficient for our purposes.

\begin{proposition}\label{generalmoments} Let $k\in \N$, and $z=re^{i\theta}$, with $(r,\theta)\in [1,\infty)\times [0,2\pi)$. Then
$$\tilde{L}(x^k,E^{z})|_{D(a^k)} = 
(Q_\theta|_{D(a)})^k+\frac {1}{r^2}C_k(r,\theta),$$
where $$C_k(r,\theta)=\sum_{\substack{n,m\in\N,\\ n+m\leq k}} c_{nm}^k(r,\theta)\, (a^*)^na^m,$$
and each function $c_{nm}^k:[1,\infty)\times [0,2\pi)\to \C$ is bounded.
\end{proposition}

\begin{proof}
Let $(r,\theta)\in [1,\infty)\times [0,2\pi)$, $z=re^{i\theta}$, and let $\psi\in \hil$, $\vp\in D(a^k)$, so that
$\vp\otimes |z\ket\in D(a^k)\otimes D(b^k)\subset D(A^k)$. By using again Proposition \ref{sqrprop},
we get $\vp\in \tilde{D}(x^k,E^z)$, and

\bet
\bra \psi |\tilde{L}(x^k, E^{re^{i\theta}})\vp\ket
= \bra \psi\otimes |z\ket |((\sqrt 2 r)^{-1} A)^k\vp\otimes |z\ket\ket
=2^{-\frac k2}r^{-k}\bra \psi\otimes |re^{i\theta}\ket |(a\otimes b^*+a^*\otimes b)^k\vp\otimes |re^{i\theta}\ket\ket.
\eeqt

The expression $(a\otimes b^*+a^*\otimes b)^k$ is the sum of all products of the form $A_1A_2\cdots A_k\otimes B_1B_2\cdots B_k$,
where for each $i=1,\ldots, k$, either $A_i=a$ and $B_i=b^*$, or $A_i=a^*$ and $B_i=b$. We can write a particular product as
$A_1A_2\cdots A_k\otimes B_1B_2\cdots B_k=p(a,a^*)\otimes p(b^*,b)$, where the symbol $p(\cdot,\cdot)$ represents the rule
telling which $A_i$:s coincide with the operator in the first argument, and which with the one in the second argument.
Let $\Pi^k$ denote the set of all such rules $p$, and let $\Pi_n^k\subset \Pi^k$ be the set of those $p$ for which
the operator in the first argument appears exactly $n$ times in the resulting product. Now $\Pi^k=\cup_{n=0}^k \Pi^k_n$, and
we have
$$(a\otimes b^*+a^*\otimes b)^k = \sum_{n=0}^k \sum_{p\in \Pi_n^k} p(a,a^*)\otimes p(b^*,b).$$
Therefore, 
$$ \bra \psi |\tilde{L}(x^k, E^{re^{i\theta}})\vp\ket = 2^{-\frac k2} \sum_{n=0}^k \sum_{p\in \Pi_n^k} \bra \psi |p(a,a^*)\vp\ket
\bra z |r^{-k}p(b^*,b) |z\ket.$$
For $p\in \Pi_n^k$, the operator $r^{-k}p(b^*,b)$, when brought to the normal order, is of the form
$r^{-k}p(b^*,b)= r^{-k}(b^*)^nb^{k-n}+r^{-k}R^{k}_p(b^*,b)$, where
$R_p^{k}$ is a polynomial of $b^*$ and $b$ of order $k-2$, with the coefficients depending on $p$. Note that in each term of
$R^k_p$, the number of occurrences of $b^*$ minus the number of occurrences of $b$ is the same, namely $n-(k-n)$.
($R_p^{k}$ is given by the Wick rule.) Hence,
$$
\bra z |r^{-k}p(b^*,b) |z\ket
= (e^{-i\theta})^n (e^{i\theta})^{k-n}(1+r^{-2}d_p^k(r)),
$$
where each function $d_p^k:[1,\infty)\to \R$ is bounded. (Of course, we could have taken any interval $[\delta,\infty)$ with
$\delta>0$.)
For $p\in \Pi_n^k$, we have $(e^{-i\theta})^n (e^{i\theta})^{k-n}p(a,a^*)=p(e^{-i\theta}a,e^{i\theta}a^*)$, so that
\beat
\bra \psi |\tilde{L}(x^k, E^{re^{i\theta}})\vp\ket &=&
2^{-\frac k2} \sum_{n=0}^k \sum_{p\in \Pi_n^k} \bra \psi |p(e^{-i\theta}a,e^{i\theta}a^*)\vp\ket\\
&+& 2^{-\frac k2} \sum_{n=0}^k \sum_{p\in \Pi_n^k}\bra \psi |r^{-2}d_p^k(r)p(e^{-i\theta}a,e^{i\theta}a^*)\vp\ket\\
&=& \left\bra \psi \big|\left([\tfrac 1 {\sqrt 2} (e^{-i\theta}a+e^{i\theta}a^*)]^k + \frac {1}{r^2} C_k(r,\theta)\right)\vp\right\ket,
\eeqat
with
$$
C_k(r,\theta) = 2^{-\frac k2} \sum_{p\in \Pi^k}d_p^k(r)p(e^{-i\theta}a,e^{i\theta}a^*).
$$
By bringing each $p(e^{-i\theta}a,e^{i\theta}a^*)$ into the normal order, and using the fact that each $d_p^k$ is a
bounded function, we see that $C_k(r,\theta)$ has the required form.
\end{proof}

The following result shows, in particular, that each observable $E^z$ is uniquely determined by its moment operator sequence
$(L(x^k,E^z))_{k\in \N}$ (see Remark \ref{determinacy}).
Hence, it is possible to consider the moment sequences in place of the observables themselves.

\begin{proposition}\label{determinacyprop} For each $z\in \C$, the semispectral measure $E^z$ is $D_{coh}$-determinate.
\end{proposition}

\begin{proof} Let $z\in \C$. Now  $D_{coh}\subset D(a^n)\subset D(x^n,E^z)$ for each $n\in \N$, so
for each $\vp\in D_{coh}$, the positive measure $E^z_\vp$ has all moments.

Consider the positive measure $E_{|\beta\ket}^z$, for fixed $\beta\in \C$.
According to \eqref{Ucoh}, we can write $U|\beta,z\ket= |c,d\ket$, with $c= \frac 1 {\sqrt 2}(\beta-z)$,
$d= \frac 1 {\sqrt 2} (\beta+ z)$.
Since $A = \frac 1 {\sqrt 2} U^*N_{-} U$, it follows from \eqref{dilation} that
$E_{|\beta\ket}^z=E^{(\sqrt 2|z|)^{-1}N_{-}}_{|c,d\ket}$. This measure
is supported in $(\sqrt 2|z|)^{-1}\Z$, with 
density $$(\sqrt 2|z|)^{-1}k\mapsto \sum_{\substack{n_1,n_2\in \N,\\ n_1-n_2 = k}} |\bra c,d|n_1,n_2\ket|^2,$$ so that
\beat
\int e^{a|x|}\,dE_{|\beta\ket}^z(x) &=& \sum_{n_1,n_2=0}^\infty e^{a(\sqrt 2|z|)^{-1}|n_1-n_2|} |\bra c|n_1\ket|^2|\bra d|n_2\ket|^2\\
&\leq& e^{-(|c|^2+|d|^2)} \sum_{n_1,n_2=0}^\infty e^{a(\sqrt 2|z|)^{-1}(n_1+n_2)} \frac {|c|^{2n_1}}{n_1!}\frac {|d|^{2n_2}}{n_2!}\\
&=& \exp [-(|c|^2+|d|^2) + e^{a(\sqrt 2|z|)^{-1}}(|c|^2+|d|^2)]\\
&=& \exp [(|\beta|^2+|z|^2)(e^{a(\sqrt 2|z|)^{-1}}-1)]<\infty.
\eeqat
Hence, $|\beta\ket \in \h E_{E^z}$ for all $\beta\in \C$.
It remains to apply Lemma \ref{exponentialboundednesslemma} to complete the proof.
\end{proof}

\section{A remark concerning characteristic functions}\label{characteristic}

Before proceeding to the conclusion of the paper, we calculate the characteristic functions of the probability
measures of the observables $E^z$, associated with the coherent states.
The purpose of this is to verify that the results are the same as those of Vogel \cite{Vogel},
obtained by a different method. Also, we want to check that the moments calculated by using these characteristic functions
are consistent with the moment operators we obtained in the preceding section.

Define $\Phi_z(t)= L(e^{it\cdot}, E^z)$ for each $t\in \R$. Since the function $x\mapsto e^{itx}$ is bounded, each $\Phi_z(t)$ is
a bounded operator. Hence, by using \eqref{dilation} and the Theorem in \cite[Section III A]{LahtiIII}, we have
\bet
\Phi(t)=V_z^*e^{it|z|^{-1} A}V_z= V_z^*U^*e^{it (\sqrt 2|z|)^{-1}N_{-}}UV_z.
\eeqt
By using \eqref{Ucoh} and putting $c= \frac 1 {\sqrt 2}(\beta-z)$,
$d= \frac 1 {\sqrt 2} (\beta+ z)$, we then get
\beat
\vp_\beta(t)&:=&\bra \beta |\Phi(t)\beta\ket\\ &=& \bra \beta, z|U^* e^{it(\sqrt 2|z|)^{-1}(\cc{I\otimes N_2-N_1\otimes I})}U |\beta,z\ket
= \bra c, d|e^{-it(\sqrt 2|z|)^{-1}N_1}\otimes e^{it(\sqrt 2 |z|)^{-1}N_2} |c, d\ket\\
&=& \bra c|e^{-it(\sqrt 2 |z|)^{-1}}c\ket\bra d|e^{it(\sqrt 2 |z|)^{-1}}d\ket
= e^{-|c|^2+\cc c e^{-it(\sqrt 2|z|)^{-1}}c} e^{-|d|^2+\cc de^{it(\sqrt 2 |z|)^{-1}}d}\\
&=&  \exp\left(-(|c|^2+|d|^2)+e^{-it(\sqrt 2 |z|)^{-1}}|c|^2+e^{it(\sqrt 2 |z|)^{-1}}|d|^2\right)\\
&=& \exp\left(-|z|^2-|\beta|^2+\tfrac 12 (\beta-z)(\cc \beta-\cc z)e^{-it(\sqrt 2|z|)^{-1}}+\tfrac 12 (\beta+z)(\cc\beta-\cc z)e^{it(\sqrt 2|z|)^{-1}}\right).
\eeqat

When replacing $z$ with $iz$ (i.e. shifting the phase by $\frac \pi 2$), and $t$ with $-t$ (accounting
for the fact that we used $I\otimes N_2-N_1\otimes I$ for the photon difference, instead of
$N_1\otimes I-I\otimes N_2$), and omitting the scale factor $\sqrt{2}$, we can compare the above result
with that of Vogel \cite[p. 17]{Vogel}, and we see that they are indeed the same.

Now $\vp_\beta(t)$ should generate the moments of the probability measure $E_{|\beta\ket}^z$ via the formula
\bet
\bra \beta |\tilde{L}(x^k,E^z)|\beta\ket = \int x^k dE_{|\beta\ket}^z = i^{-k} \frac{d^k\vp_\beta}{dt^k}(0).
\eeqt

We check the first two moments for consistency. Let again $z=re^{i\theta}$. Clearly, $\vp_\beta(0)=1$. Differentiating $\vp_\beta$ gives

\bet
\vp_\beta'(t) = i\tfrac 1 {\sqrt 2}|z|^{-1}(-|c|^2e^{-i(\sqrt 2|z|)^{-1}t}+|d|^2e^{i(\sqrt 2|z|)^{-1}t})\vp_\beta(t),
\eeqt
so that $\vp_\beta'(0) =  i\tfrac 1 {\sqrt 2} |z|^{-1}(-|c|^2+|d|^2) = \tfrac 1 {\sqrt 2} i|z|^{-1}(z \cc\beta+\cc z\beta)$, and hence
the first moment of the probability measure $E^z_{|\beta\ket}$ is
\bet
-i\vp_\beta'(0)= \tfrac 1 {\sqrt 2} (e^{-i\theta} \beta +e^{i\theta}\cc\beta)
= \bra \beta |\tfrac 1 {\sqrt 2}(e^{-i\theta}a+e^{i\theta} a^*)|\beta\ket,
\eeqt
in agreement with Proposition \ref{fstsndmoments} (a).

In addition,
\bet
\vp_\beta''(t) = -\tfrac 12|z|^{-2} \left((|c|^2e^{-i(\sqrt 2 |z|)^{-1}t}+|d|^2e^{i(\sqrt 2 |z|)^{-1}t}) +
(-|c|^2e^{-i(\sqrt 2 |z|)^{-1}t}+|d|^2e^{i(\sqrt 2 |z|)^{-1}t})^2\right)\vp_\beta(t),
\eeqt
so the second moment is
\beat
i^{-2}\vp_\beta''(0) &=& \tfrac 12 |z|^{-2}(|\beta|^2+|z|^2) + \tfrac 12 |z|^{-2} [(\cc z \beta +\cc \beta z)]^2 \\
&=& \tfrac 12 ([(e^{-i\theta}\beta +e^{i\theta} \cc\beta)]^2 +1) + \tfrac 12|\beta|^2r^{-2}\\
&=& \tfrac 12(e^{-2i\theta}\beta^2+e^{2i\theta}\cc \beta^2+(|\beta|^2+1)+|\beta|^2) +\tfrac 12|\beta|^2 r^{-2}\\
&=& \bra \beta |\tfrac 12(e^{-2i\theta} a^2+e^{2i\theta}(a^*)^2+aa^*+a^*a)|\beta\ket + \bra \beta| \tfrac 12r^{-2} a^*a|\beta\ket\\
&=& \bra \beta |([\tfrac 1 {\sqrt 2}(e^{-i\theta}a+e^{i\theta} a^*)]^2+\tfrac 12 r^{-2}a^*a)|\beta\ket,
\eeqat
which agrees with the equation of Proposition \ref{fstsndmoments} (b).

\section{Convergence in the high amplitude limit}\label{conclusion}

We are now ready to establish a conclusion on the high amplitude limit, by connecting the
results concerning the detector observables $E^z$ with the ''asymptotic measurement'' procedure described at the end
of Section \ref{generalsection}.
Since the considerations of that section involved sequences of semispectral measures, we need to fix
a sequence $(r_n)$ of positive numbers converging to infinity. For this choice, let $z_n(\theta) = r_n e^{i\theta}$,
where the phase $\theta\in [0,2\pi)$ is also fixed.

Choose the set $D_{coh}$ of coherent states to be the subspace containing the calibration states mentioned in part
\eqref{item1} of the aforementioned procedure. 
It follows immediately from Proposition \ref{generalmoments} that for each $\theta\in [0,2\pi)$,
the spectral measure of the rotated quadrature $Q_\theta$
is a moment limit for $(E^{z_n(\theta)})_{n\in \N}$ on $D_{coh}$, so that the requirements for parts \eqref{item2} and \eqref{item3}
are satisfied. The determinacy requirement of part \eqref{item4} is given by lemma \ref{quadraturedeterminacy},
so that Proposition \ref{momentasymptotic} (b) implies that the sequence $(E^{z_n(\theta)})_{n\in \N}$ has only one
moment limit on $D_{coh}$, namely $E^{Q_\theta}$.

Finally, the requirement for part \eqref{item5} is satisfied by proposition \ref{determinacyprop}. Thus, we can conclude that
the (above simple theoretical description of the) balanced homodyne detector fits into our ''asymptotic measurement'' scheme.
In particular, the sequence $(E^{z_n(\theta)})_{n\in \N}$ of observables converges to the spectral measure of
the rotated quadrature $Q_\theta$ weakly in the sense of probabilities.

\begin{remark} \rm By definition, the weak convergence in the sense of probabilities of $(E^{z_n(\theta)})_{n\in \N}$ to $E^{Q_\theta}$
means that the sequence $(E^{z_n(\theta)}(B))_{n\in \N}$ converges to $E^{Q_\theta}(B)$ in the weak operator topology,
provided that $E^{Q_\theta}(\partial B)=0$. Since $Q_\theta$ is unitarily equivalent to the position operator on $L^2(\R)$,
this condition is equivalent to $\lambda(\partial B) = 0$, where $\lambda$ is the Lebesgue measure
of the real line. In particular, all intervals (finite or infinite) qualify.
\end{remark}
\begin{remark}\rm
It should be emphasized that $\lim_{n\goto\infty} E^{z_n(\theta)}(B) = E^{Q_\theta}(B)$
(in the weak operator topology) is \emph{not} true for all $B\in \h B(\R)$.
In fact, let $B = \bigcup_{n\in \N}(\sqrt 2 r_n)^{-1}\Z$, where $z_n(\theta) = r_ne^{i\theta}$ as before.
Now $B$ is countable, and, in particular, $B\in \h B(\R)$.
It follows from \eqref{dilation} and the relation $A = \frac 1 {\sqrt 2}U^*N_-U$ that that each
semispectral measure $E^{z_n(\theta)}$ is supported on $(\sqrt 2 r_n)^{-1}\Z$. Hence,
$E^{z_n(\theta)}(B)=I$ for all $n\in \N$, so that $(E^{z_n(\theta)}(B))_{n\in \N}$ converges to $I$ in the weak
operator topology. But $E^{Q_\theta}(B) = O$, because $B$ is countable and
$E^{Q_\theta}$ is unitarily equivalent to the position operator on $L^2(\R)$, thereby having the same null sets as the
Lebesgue measure. On the other hand, $B$ is dense in $\R$, so $\partial B = \R$, and consequently $E^{Q_\theta}(\partial B)=I$, which is not zero. 
\end{remark}

The above conclusions are summarized as follows.

\begin{itemize}
\item[(a)] The observable $E^{z}$ measured by the balanced homodyne detector is $D_{coh}$-determinate for each $z\in \C$;
\item[(b)] The spectral measure $E^{Q_\theta}$ of the rotated quadrature $Q_\theta$ is $D_{coh}$-determinate for each $\theta\in [0,2\pi)$.
\item[(c)] For each $\theta\in [0,2\pi)$, and each discretization $z_n(\theta) = r_n e^{i\theta}$, $\lim_{n\goto\infty} r_n = \infty$,
the spectral measure $E^{Q_\theta}$ is the unique moment limit of $(E^{z_n(\theta)})_{n\in \N}$ on $D_{coh}$.
\item[(d)] The sequence $(E^{z_n(\theta)})_{n\in \N}$ converges to $E^{Q_\theta}$ weakly in the sense of probabilities.
\item[(e)] For each state $T\in \h S(\hil)$, we have $\lim_{n\goto\infty}\tr[TE^{z_n(\theta)}(B)] = \tr[TE^{Q_\theta}(B)]$,
whenever $\lambda(\partial B)=0$, where $\lambda$ is the Lebesque measure on $\R$.  
\end{itemize}

\section*{Appendix A}

In this appendix, we give a proof for proposition \ref{weakconvergence}.
Since it does not depend on the special properties of the real line, we state it here
in the context of a general metric space. For the rest of this section, we fix $\Omega$ to be a metric space, with metric $d$, and
let $\h B(\Omega)$ denote the associated Borel $\sigma$-algebra. As before, $\hil$ is a complex separable Hilbert space,
and the general notations given in section \ref{notations} will be used.

We recall the definition of the weak convergence of probability measures \cite[p. 11]{Billingsley}:
a sequence $(\mu_n)$ of probability measures on $\h B(\Omega)$ converges weakly to a probability measure
$\mu:\h B(\Omega)\to [0,1]$ if $\lim_{n\goto\infty} \int f \,d\mu_n =\int f\,d\mu$ for all bounded continuous functions $f:\Omega\to \R$.
We will use the following characterization for the weak convergence \cite[Theorem 2.1, p. 11]{Billingsley}:
sequence $(\mu_n)$ of probability measures on $\h B(\Omega)$ converges weakly to a probability measure
$\mu:\h B(\Omega)\to [0,1]$ if and only if $\lim_{n\goto\infty} \mu_n(A)= \mu(A)$ whenever $A\in \h B(\Omega)$ is such that $\mu(\partial A)=0$.
At the level of semispectral measures $E:\h B(\Omega)\to L(\hil)$, a natural analogue for the weak convergence is the following.

\begin{definition}\label{weakdef2}\rm Let $E^n:\h B(\Omega)\to L(\hil)$ be a semispectral measure for each $n\in \N$. We say that the sequence
$(E^n)$ converges to a semispectral measure $E:\h B(\Omega)\to L(\hil)$ \emph{weakly in the sense of probabilities}, if
$$ \lim_{n\goto\infty} E^n(A) = E(A)$$ in the weak operator topology, for all $A\in \h B(\Omega)$ such that $E(\partial A)=0$.
\end{definition}

The next proposition characterizes this convergence in terms of the weak convergence of probability measures.
First we need the following lemma.
\begin{lemma}\label{algebralemma} Let $\h P$ be an at most countable collection of finite positive measures $\nu:\h B(\Omega)\to [0,\infty)$.
Define $\h F_{\h P} = \{ A\in \h B(\Omega)\mid \nu(\partial A) = 0 \text{ for all } \nu\in \h P\}$.
\begin{itemize}
\item[(a)] $\h F_{\h P}$ is an algebra which generates the Borel $\sigma$-algebra $\h B(\Omega)$.
\item[(b)] If $(\mu_n)$ is a sequence of probability measures on $\h B(\Omega)$, such that $\lim_{n\goto\infty}\mu_n(A) = \mu(A)$
for all $A\in \h F_{\h P}$, where $\mu:\h B(\Omega)\to [0,1]$ is also a probability measure, then $(\mu_n)$ converges to $\mu$ weakly.
\end{itemize}
\end{lemma}
\begin{proof} Since $\partial (A\cup B)\subset \partial A\cup \partial B$, and $\partial A = \partial (\Omega\setminus A)$ for any $A, B\subset \Omega$, it is clear that $\h F_{\h P}$ is an algebra.
The rest of the proof uses an argument similar to one appearing in the proof of \cite[Theorem 4]{Ressel}:
Let $G\subset \Omega$ be an open set. For each $\delta>0$, let $A_\delta = \{x\in \Omega\mid d(x,\Omega\setminus G) >\delta\}$.
Now $A_\delta$ is open, and hence a Borel set.
Since $\partial A_\delta\subset \{ x\in \Omega\mid d(x,\Omega\setminus G) = \delta\}$, the sets $\partial A_\delta$
are disjoint for distinct values of $\delta$. Let $\nu\in \h P$. Since $\nu$ is a finite positive measure, the family
$\{ \nu(\partial A_\delta)\mid \delta >0\}$ is summable, and hence the set
$\{\delta >0\mid \nu(\partial A_\delta)>0\}$ is at most countable. Since $\h P$ is at most countable by assumption,
also the set $\{\delta >0\mid A_\delta \notin \h F_{\h P}\}$ has this property. This implies that there exists a sequence
$(\delta_n)$ of positive numbers converging to $0$, with $A_{\delta_n}\in \h F_{\h P}$ for all $n\in \N$.
But clearly $G = \bigcup_{n\in \N} A_{\delta_n}$, so $\h F_{\h P}$ generates $\h B(\Omega)$. This proves (a), and
(b) follows from \cite[Theorem 2.2, p. 14]{Billingsley}.
\end{proof}

Some parts of the proof of the following proposition can be extracted at least from \cite{Blum} or \cite{Ressel}
(the latter concerning only multiplicative operator measures),
so one might expect that the result is essentially well-known. However, having not been able to find the entire proof in the
literature, we give it here.

\begin{proposition}\label{weakconvergence2} Let
$E^n:\h B(\Omega)\to L(\hil)$ be a semispectral measure for each $n\in \N$, and let also $E:\h B(\Omega)\to L(\hil)$ be a semispectral measure.
Then the following conditions are equivalent.
\begin{itemize}
\item[(i)] $(E^n)$ converges to $E$ weakly in the sense of probabilities;
\item[(ii)] for each positive operator $T$ of trace one, the sequence $(E^n_T)$ of probability measures converges
weakly to $E_T$;
\item[(iii)] there exists a dense subspace $\h D\subset \hil$, such that the sequence $(E^n_\vp)$ of
probability measures converges weakly to $E_\vp$ for any unit vector $\vp\in \h D$;
\item[(iv)] $\lim_{n\goesto \infty} L(f,E^n) = L(f,E)$ in the weak operator topology for each bounded continuous
function $f:\Omega\to \R$.
\end{itemize}
\end{proposition}
\begin{proof}
To prove that (i) implies (ii), we assume (i), so that the sequence $(\bra \vp| E^n(A)\vp\ket)_{n\in \N}$ of numbers
converges to $\bra \vp |E(A)\vp\ket$ for each $\vp\in \hil$ and $A\in \h G$ where
$\h G = \{ A\in \h B(\Omega)\mid E(\partial A)=0\}$. Since $E$ is a semispectral measure and $\hil$ is separable,
there exists a finite positive measure $\nu:\h B(\Omega)\to [0,\infty)$ with the same sets of measure zero as
$E$. (One can take, for instance, $\nu(A) = \tr[SE(A)]$, with $S= \sum_{n\in \N} \tfrac 1 {2^n} |\eta_n\ket\bra\eta_n|$,
where $\{\eta_n\mid n\in \N\}$ is an orthonormal basis of $\hil$.) Hence, using the notation of Lemma \ref{algebralemma}, we have $\h G= \h F_{\{\nu\}}$.

Now fix a positive operator $T\in L(\hil)$
with unit trace, and let $A\in \h G$. Write $T=\sum_{k=1}^\infty t_k |\vp_k\ket\bra\vp_k|$, where the $\vp_k$ are unit vectors and
$\sum_n t_n =1$, the series converging in the trace norm. Now we have
\begin{eqnarray*}
\lim_{n\goto\infty} \tr[TE^n(A)]&=& \lim_{n\goto\infty} \sum_{k=1}^\infty t_k \bra \vp_k |E^n(A)\vp_k\ket
= \sum_{k=1}^\infty t_k \lim_{n\goto\infty} \bra \vp_k |E^n(A)\vp_k\ket\\
&=& \sum_{k=1}^\infty t_k \bra \vp_k |E(A)\vp_k\ket = \tr[TE(A)],
\end{eqnarray*}
where the change of the order of the limit procedures is permissible, since the inequality $\|E^n(A)\|\leq 1$ implies
that the series
$$\sum_{k=1}^\infty t_k \bra \vp_k |E^n(A)\vp_k\ket$$ converges \emph{uniformly} for $n\in \N$.
Hence, $(E^n_T(A))$ converges to $E_T(A)$ for all $A\in \h G$. Now Lemma \ref{algebralemma} (b) gives (ii).

Clearly (ii) implies (iii). Now we assume (iii) and prove that (i) holds. Let $\h D$ be the dense subspace in question,
and $A\in \h B(\Omega)$ be such that
$E(\partial A)=0$. Then $E_\vp(\partial A)=0$ for all unit vectors $\vp\in \h D$, so that
$\lim_{n\goto\infty} E^n_\vp(A) = E_\vp(A)$ for all $\vp\in \h D$, $\|\vp\|=1$ by (iii). Let $\vp\in \hil$, and
select a sequence $(\vp_k)$ converging to $\vp$, such that $\vp_k\in \h D$ for all $k$. Then
$$
\lim_{n\goto\infty}\bra \vp|E^n(A)\vp\ket =\lim_{n\goto\infty}\lim_{k\goto\infty}\bra \vp_k|E^n(A)\vp_k\ket
= \lim_{k\goto\infty}\lim_{n\goto\infty}\bra \vp_k|E^n(A)\vp_k\ket = \bra \vp|E(A)\vp\ket,$$
where the change of the order of the limit procedures is justified since the $k$ limit is uniform in $n$.
Hence, (i) follows by polarization.

Finally, it is clear from \cite[Theorem 2.1]{Billingsley} that (ii) implies (iv) and (iv) implies (iii),
since e.g. $\bra \vp|L(f,E)\vp\ket = \int f dE_{\vp}$ for any unit vector $\vp\in \hil$, and a bounded continuous
function $f:\Omega\to \R$. (Notice that e.g. $L(f,E)\in L(\hil)$ because $f$ is bounded.) The proof is complete.

\end{proof}

\begin{remark}\rm Since a sequence of probability measures cannot converge weakly to two different limits
\cite[Theorem 1.3, p. 9]{Billingsley}, the above proposition shows, in particular, that a sequence of operator measures
can converge to at most one operator measure weakly in the sense of probabilities.
\end{remark}

\section*{Appendix B}

This appendix is devoted to the proof of Proposition \ref{weaklycompact2}. We use the same general context
of a metric space as in Appendix A, as well as the definitions and notations given there.

In the context of probability theory, a family $\h P$ of probability measures $\nu: \h B(\Omega)\to [0,1]$ is called
\emph{relatively compact}, if every sequence of elements of $\h P$ contains a weakly convergent subsequence
(see \cite[p. 35]{Billingsley}). 

It is an old result that for a sequence of probability measures on $\R$, the convergence of moments (which was a central concept in
the considerations of section \ref{generalsection}) and weak convergence
are related, in the case where the limiting measure is determinate (see \cite[p. 405-408]{BillingsleyII},
the original paper \cite{Frechet} by Fr\'echet and Shohat, and also \cite{Rao}). This is connected with the fact that
the relevant sequences of probability measures are relatively compact
(see e.g. \cite[Theorems 6.1 and 6.2, p. 37]{Billingsley} and the proof of \cite[Theorem 30.2, p. 408]{BillingsleyII}).
The following result shows that the relative compactness of probability measures is reflected in a natural way in
the level of operator measures.

\begin{proposition}\label{weaklycompact} Let $\h D\subset\hil$ be a dense subspace,
and let $\h M$ be a collection of semispectral measures $E:\h B(\Omega)\to L(\hil)$.
Suppose that the set $\{E_{\vp}\mid E\in \h M\}$ of probability measures is relatively compact for each unit vector
$\vp\in \h D$.
Then every sequence of elements of $\h M$ contains a subsequence which converges weakly in the sense of probabilities.
\end{proposition}
\begin{proof}

Let $\{\vp_n\mid n\in \N\}$ be an orthonormal basis of $\hil$, included in $\h D$. (Since $\hil$ is separable
and $\h D$ is dense, such a basis exists.) Let $\h V$ be the linear span of this basis. Then let
$$\h K = \{ \|i^l\vp_n +\vp_m\|^{-1}(i^l\vp_n+\vp_m)\mid n,m\in \N, \, l=0,1,2,3\}.$$
Now $\h K$ is countable, so we can put $\h K = \{\psi_j\mid j\in \N\}$, where $\psi_i\neq \psi_j$ for $i\neq j$.

Let $(E^n)$ be a sequence of elements of $\h M$. For each $j\in \N$ and $n\in \N$, let $E^n_j$ denote the probability measure $E^n_{\psi_j}$.
Since $\psi_j\in \h D$, the sequence $(E^n_j)_{n\in \N}$ (and every one of it subsequences) contains a weakly convergent subsequence
by assumption.

Now we can use the classic ''diagonal process'' to get a subsequence
$(E^{n_k})$ of $(E^n)$, and a sequence of probability measures $(\mu_j)$, such that
$(E^{n_k}_j)_{k\in \N}$ converges to $\mu_j$ weakly for all $j\in \N$.


The rest of the proof uses somewhat similar arguments to the proof of \cite[Theorem 1]{Blum}.

For each $i,j\in \N$, $l=0,1,2,3$, define
$\tilde{\mu}^l_{ij}:=\mu_n$, with $n\in\N$ (the unique index) such that $\psi_n = \|i^l\vp_i+\vp_j\|^{-1}(i^l\vp_i+\vp_j)$.
Then for each $i,j\in \N$, define
$$\nu_{ij}(A) = \frac 14 \sum_{l=0}^3 i^l \|i^l\vp_i+\vp_j\|^2\tilde{\mu}_{ij}^l(A), \ \ \ A\in \h B(\Omega).$$
Now each $\nu_{ij}$ is clearly a complex measure, with the property that $\nu_{ij}(\Omega) = \bra \vp_i|\vp_j\ket$ for any $i,j$.
Finally, for each $\vp,\psi\in \h V$, set
\be\label{complexmeasure}
G(A)(\psi,\vp) = \sum_{i,j\in \N} \bra\psi|\vp_i\ket\bra\vp_j|\vp\ket \nu_{ij}(A), \ \ \ A\in \h B(\Omega), \vp,\psi\in \h V.
\eeq
Note that the above double sum is always finite, since $\vp,\psi\in \h V$. Hence, the set function
$\h B(\Omega)\ni A\mapsto G(A)(\psi,\vp)\in \C$ is a complex measure, which satisfies
\be\label{normalization}
G(\Omega)(\psi,\vp) = \bra \psi|\vp\ket, \ \ \ \psi,\vp\in \h V.
\eeq
For each $A\in \h B(\Omega)$, we now have a map $G(A):\h V\times \h V\to \C$, which is clearly sesquilinear by its definition
\eqref{complexmeasure}. We have to prove that it is positive, i.e.
\be\label{positivity}
G(A)(\vp,\vp)\geq 0, \ \ \ \vp\in \h V, A\in \h B(\Omega).
\eeq
To that end, let $\h P = \{ \mu_j \mid j\in \N\}$, so that
$$\lim_{k\goto\infty} \bra \psi_j|E^{n_k}(A)\psi_j\ket =\mu_j(A), \ \ \  j\in \N, \,A\in \h F_{\h P},$$
with $\h F_{\h P}$ as in Lemma \ref{algebralemma}. This means that
$$\lim_{k\goto\infty} \bra i^l\vp_i+\vp_j|E^{n_k}(A)(i^l\vp_i+\vp_j)\ket = \|i^l\vp_i+\vp_j\|^2\tilde{\mu}_{ij}^l(A), \  \ \ i,j\in \N,\, l=0,1,2,3, \,A\in \h F_{\h P}.$$
It now follows from \eqref{complexmeasure} (by using the polarization identity to the sesquilinear forms
$(\psi,\vp)\mapsto \bra\psi|E^{n_k}(A)\vp\ket$) that
\be\label{converg}
\lim_{k\goto\infty} \bra \psi|E^{n_k}(A)\vp\ket = G(A)(\psi,\vp), \ \ \ A\in \h F_{\h P},\, \vp,\psi\in\h V.
\eeq
Since each $E^{n_k}(A)$ satisfies $0\leq \bra \vp |E^{n_k}(A)\vp\ket$ for all $\vp\in \h V$,
this clearly implies that the same is true for $G(A)$, provided $A\in \h F_{\h P}$.
We have to prove that this is true for all $A\in \h B(\Omega)$. To that end, let
$$
\h B_1 = \{ A\in \h B(\Omega) \mid 0\leq G(A)(\vp,\vp) \text{ for all } \vp\in \h V\}.
$$
Now $\h F_{\h P}\subset \h B_1$. We show that $\h B_1$ is a monotone class. Let $(A_n)$ be an increasing sequence
of sets in $\h B_1$, and let $A=\cup_{n=1}^\infty A_n$. Let $\vp\in \h V$. Since $B\mapsto G(B)(\vp,\vp)$ is a complex measure,
we get $$G(A)(\vp,\vp) = \lim_{n\goto\infty} G(A_n)(\vp,\vp)\geq 0,$$
so that $\cup_{n=1}^\infty A_n=A\in \h B_1$.
Similarly, we see that $\cap_{n=1}^\infty A_n\in \h B_1$ for a decreasing sequence $(A_n)$ of sets of $\h B_1$.
This shows that $\h B_1$ is a monotone class. Since $\h F_{\h P}\subset \h B_1$, it follows from the monotone class theore that $\h B_1$ contains the $\sigma$-algebra generated by $\h F_{\h P}$. But this is $\h B(\Omega)$ by
Lemma \ref{algebralemma} (a), so that \eqref{positivity} holds.

Now \eqref{positivity} and \eqref{normalization} imply that 
$0\leq G(A)(\vp,\vp)\leq \|\vp\|^2$ for any $\vp\in \h V$ and $A\in \h B(\Omega)$. By polarization, this implies that
$$\sup \{|G(A)(\vp,\psi)|\mid A\in \h B(\Omega), \vp,\psi\in \h V, \|\vp\|\leq 1, \|\psi\|\leq 1\}<\infty.$$
Since $\h V$ is dense, it follows that for each $A\in \h B(\Omega)$, there exists a unique operator $E(A)\in L(\hil)$, such that
$G(A)(\vp,\psi) = \bra \vp|E(A)\psi\ket$ for all $\vp,\psi\in \h V$. Now \eqref{positivity} and \eqref{normalization}
imply that 
the map $E:\h B(\Omega)\to L(\hil)$ is a semispectral measure.

By \eqref{converg}, $\lim_{k\goto \infty}E^{n_k}_\vp(A)= E_\vp(A)$ for all $A\in \h F_{\h P}$ and each unit vector $\vp\in \h V$. It follows from Lemma \ref{algebralemma} (b) that for each unit vector $\vp\in \h V$, the sequence $(E^{n_k}_\vp)_{k\in\N}$ of
probability measures converges weakly to the probability measure $E_\vp$. According to Proposition \ref{weakconvergence2}, this
means that $(E^{n_k})_{k\in \N}$ converges to $E$ weakly in the sense of probabilities.
\end{proof}

\noindent {\bf Acknowledgments.} The authors thank Dr. Paul Busch and Dr. Kari Ylinen for useful discussions and
comments. One of us (J.K.) was supported by Finnish Cultural Foundation.

\end{document}